\documentclass{svproc}
\usepackage[T1]{fontenc}
\usepackage[utf8]{inputenc}

\usepackage{color}
\usepackage{multirow}
\usepackage{amssymb}
\usepackage{graphicx}
\usepackage{enumitem}
\usepackage{indentfirst}
\usepackage{caption}
\usepackage{verbatim}

\usepackage[numbers,square]{natbib}
\makeatletter
\renewcommand\@biblabel[1]{#1.}
\makeatother

\usepackage{url}

\begin{document}
\mainmatter              
\title{Metadata Systems for Data Lakes:\\ Models and Features}
\titlerunning{Metadata Systems for Data Lakes}  
%
\author{Pegdwendé N. Sawadogo\inst{1} \and \'Etienne Scholly\inst{1,2} \and \\
Cécile Favre\inst{1} \and \'Eric Ferey\inst{2} \and Sabine Loudcher\inst{1} \and Jérôme Darmont\inst{1}}
\authorrunning{P.N. Sawadogo et al.} 
%
%
\institute{Université de Lyon, Lyon 2, ERIC EA 3083\\
\url{https://eric.ish-lyon.cnrs.fr/}
\and
BIAL-X\\
\url{https://www.bial-x.com/}
}

\maketitle              

\begin{abstract} 
    Over the past decade, the data lake concept has emerged as an alternative to data warehouses for storing and analyzing big data. A data lake allows storing data without any predefined schema. Therefore, data querying and analysis depend on a metadata system that must be efficient and comprehensive. However, metadata management in data lakes remains a current issue and the criteria for evaluating its effectiveness are more or less nonexistent.
    
    In this paper, we introduce MEDAL, a generic, graph-based model for metadata management in data lakes. We also propose evaluation criteria for data lake metadata systems  through a list of expected features. Eventually, we show that our approach is more comprehensive than existing metadata systems.
    
\end{abstract}

\section{Introduction}

    Since the beginning of the 21\textsuperscript{st} century, the usages of organizations in decision-making processes have been disrupted by the availability of large amounts of data, i.e., big data. Mainly issued from social media and the Internet of things, big data bring about great opportunities for organizations, but also issues related to data volume, velocity and variety, which surpass the capabilities of traditional storage and data processing systems \cite{Miloslavskaya2016}. 
    
    In this context, the concept of data lake  \cite{Dixon2010} appears as a solution to big data heterogeneity problems. A data lake provides integrated data storage without predefined schema~\cite{Hai2016}. In the absence of a data schema, an effective metadata system becomes essential to make data queryable and thus prevent the lake from turning into a data swamp, i.e., an inexploitable data lake~\cite{Alrehamy2015,Hai2016,Suriarachchi2016}.
    
    While the literature seems unanimous about the importance of the metadata system in a data lake, questions and uncertainties remain about its implementation methodology. Several approaches help organize metadata, but most concern only structured and semi-structured data~\cite{Farid2016, Hai2016, Maccioni2018, Quix2016B}. Moreover, the effectiveness of a metadata system is difficult to measure because, to the best of our knowledge, there are no widely shared and accepted evaluation criteria.
    
    To address these issues, we first identify a set of features that should ideally be proposed by the metadata system of a data lake. By comparing several metadata systems with respect to these features, we hint that none of them offers all expected features. Thus, we propose a new metadata model that is more complete and generic. Our graph-based metadata model is named MEtadata model for DAta Lakes (MEDAL). It is also based on a  typology distinguishing intra-object, inter-object and global metadata.
    
    The remainder of this paper is organized as follows. Section~\ref{sec:definitions} introduces the concept of data lake. Section~\ref{sec:litterature} details the expected features of a data lake's metadata system and compares several works on the organization of metadata w.r.t. these features. Section~\ref{sec:typ_metadata} presents the metadata typology on which MEDAL is based. Section~\ref{sec:repres_concep} formalizes MEDAL and introduces its graph representation. Finally, Section~\ref{sec:conclusion} concludes the paper and hint at research perspectives.

\section{Data Lake Concept}
\label{sec:definitions}
    
    \subsection{Definitions from the Literature}
    \label{sec:def_litterature}

        James Dixon introduces the data lake concept as an alternative to data marts, which are subsets of data warehouses that store data into silos. A data lake is a large repository of heterogeneous raw data, supplied by external data sources and from which various analyses can be performed \cite{Dixon2010}.
        
        Thereafter, data lakes are associated with the Hadoop technology \cite{Fang2015,OLeary2014}. Data lake design may notably be viewed as a methodology for using free or low-cost technologies, typically Hadoop, for storing, processing and exploring raw data within a company \cite{Fang2015}. However, this view is becoming minority in the literature, as the data lake concept is now also associated with proprietary solutions such as Azure or IBM~\cite{Madera2016, Sirosh2016}.
        
        A more consensual definition is to see a data lake as a central repository where data of all formats are stored without a strict schema~\cite{Khine2017, Laskowski2016, Mathis2017}. This definition is based on two key characteristics of  data lakes:  data variety and the schema-on-read (or late binding) approach, which consists in defining the data schema at analysis time \cite{Miloslavskaya2016}.  
        
        However, the variety/schema-on-read definition provides little detail about the characteristics of a data lake. Thus, a more complete definition by Madera and Laurent views data lake as a logical view of all data sources and datasets in their raw format, accessible by data scientists or statisticians for knowledge extraction \cite{Madera2016}. This definition is complemented by a list of key features: 1)~data quality  is provided by a set of metadata; 2)~the lake is controlled by data governance policy tools; 3)~usage of the lake is limited to statisticians and data scientists; 4)~the lake integrates data of all types and formats; 5)~the data lake has a logical and physical organization.

    \subsection{Discussion and New Definition}
    \label{sec:our_def}
        
        Some points in Madera and Laurent's definition of data lakes  \cite{Madera2016} are debatable. The authors indeed reserve the use of the lake to data specialists and, as a consequence, exclude business experts for security reasons. Yet, in our opinion, it is entirely possible to allow controlled access to this type of users through a navigation or analysis platform.
        
        Moreover, we do not share the vision of the data lake as a logical view over data sources, since some data sources may be external to an organization, and therefore to the data lake. Dixon also specifies that lake data come from data sources \cite{Dixon2010}. Including data sources into the lake may therefore be considered contrary to the spirit of data lakes.
        
        Finally, although quite complete, Madera and Laurent's definition omits an essential property of data lakes: 
        scalability \cite{Miloslavskaya2016}. Since a data lake is intended for big data storage and processing, it is indeed essential to address this issue.
        
        Thence, we amend Madera and Laurent's definition to bring it in line with our vision and introduce scalability.
      
        \begin{definition}
            A data lake is a scalable storage and analysis system for data of any type, retained in their native format and used \emph{mainly} by data specialists (statisticians, data scientists or analysts) for knowledge extraction. Its characteristics include: 1)~a metadata catalog that enforces data quality; 2)~data governance policies and tools; 3)~accessibility to various kinds of users; 4)~integration of any type of data; 5)~a logical and physical organization; 6)~scalability.
        \end{definition}

\section{Basic Features of a Metadata System}
\label{sec:litterature}

    \subsection{Expected Features}
    \label{sec:fonct}
    
        We identify in the literature six main functionalities that should ideally be provided by the metadata system of a data lake.
        
        \noindent\textbf{Semantic enrichment (SE)}, also called semantic annotation~\cite{Hai2016} or semantic profiling~\cite{Ansari2018}, consists in generating a description of the context of data, e.g., with tags, to make them more interpretable and understandable~\cite{Terrizzano2015}. It is done using knowledge bases such as ontologies.
        Semantic annotation plays a key role in data exploitation, by summarizing the datasets contained in the lake so that they are more understandable to the user. It can also be used as a basis for identifying data links. For instance, data associated with the same tags can be considered linked.
        
        \noindent\textbf{Data indexing (DI)} consists in setting up a data structure to retrieve datasets based on specific characteristics (keywords or patterns). 
        This requires the construction of forward or inverted indexes.
        Indexing makes it possible to optimize data querying in the lake through keyword filtering. It is particularly useful for textual data management, but can also be used in a semi-structured or structured data context~\cite{Singh2016}. 
        
        \noindent\textbf{Link generation and conservation (LG)} is the process of detecting similarity relationships or integrating preexisting links between datasets.
        The integration of data links can be used to expand the range of possible analyses from the lake by recommending data related to those of interest to the user~\cite{Maccioni2018}. Data links can also be used to identify data clusters, i.e., groups where data are strongly linked to each other and significantly different from other data~\cite{Farrugia2016}.

        \noindent We define \textbf{data polymorphism (DP)} as storing multiple representations of the same data. 
        Each representation corresponds to the initial data, modified or reformatted for a specific need. For example, a textual document can be represented without stopwords or as a bag of words.
        It is essential in the context of data lakes to at least partially structure unstructured data to allow their automated analysis~\cite{Diamantini2018}. 
        Simultaneously storing several representations of the same data notably avoids repeating preprocessings and thus speeds up analyses.
      
        \noindent\textbf{Data versioning (DV)} refers to the ability of the metadata system to support data changes while conserving previous states. 
        This ability is essential in data lakes, as it ensures the reproducibility of analyses and supports the detection and correction of possible errors or inconsistencies. Versioning also allows to support a branched evolution of data, especially in their schema~\cite{Hellerstein2017}.
        
        \noindent\textbf{Usage tracking (UT)} records the interactions between users and the data lake. Interactions are generally operations of data creation, update and access. 
        The integration of this information into the metadata system makes it possible to understand and explain possible inconsistencies in the data~\cite{Beheshti2017}. It can also be used to manage sensitive data, by detecting intrusions~\cite{Suriarachchi2016}. 
        
        Usage tracking and data versioning are closely linked, because interactions lead  in some cases to the creation of new versions or representations of the data. Thus, such features are often integrated together in a provenance tracking module~\cite{Halevy2016,Hellerstein2017,Terrizzano2015}. Yet, we still consider that they remain different features since they are not systematically proposed together~\cite{Beheshti2017, Diamantini2018, Suriarachchi2016}.

    \subsection{Comparison of Metadata Systems}
    \label{sec:comparaison_fonct}

        We consider in this comparison two types of metadata systems: metadata models and data lake implementations. 
        Metadata models refer to conceptual systems for organizing metadata. They have the advantage of being more detailed and more easily reproducible than data lake implementations, which lie at a more operational level. 
        Data lake implementations focus on operation and functionality, with little detail on the conceptual organization of metadata.
        Eventually, we include in this study systems (models or implementations) not explicitly associated with the concept of data lake by their authors, but that may be used in a data lake context, e.g., the Ground metadata model~\cite{Hellerstein2017}.
        
        Table~\ref{tab:comparaison} provides a synthetic comparison of 15 metadata systems for data lakes (and assimilated). It shows that the most complete systems in terms of functionality are the GOODS and CoreKG data lakes, with five out of six features available. These systems notably support polymorphism and data versioning. However, they are also black boxes providing little detail on the conceptual organization of metadata. Ground may therefore be preferred, since it is much more detailed and almost as complete (4/6).
        
        \begin{table*}[ht]
            \centering
              \caption{Features provided by data lake metadata systems}
            \begin{tabular}{r c c l l l l l l l}
            \hline 
                \textbf{System} &  & \hspace*{-0.5cm}\textbf{Type} & \textbf{SE} & \textbf{DI} & \textbf{LG} & \textbf{DP}  & \textbf{DV} & \textbf{UT}\\  \hline  
                
                SPAR (Fauduet and Peyrard, 2010)&\citep{Fauduet2010} & \multirow{1}{0.7cm}{$\blacklozenge \sharp$} & \multirow{1}{0.7cm}{\checkmark} & \multirow{1}{0.7cm}{\checkmark} & \multirow{1}{0.7cm}{\checkmark} & \multirow{1}{0.7cm}{} & \multirow{1}{0.7cm}{} & \multirow{1}{0.7cm}{\checkmark}\\ \hline
                
                Alrehamy and Walker (2015)&\cite{Alrehamy2015}& \multirow{1}{0.7cm}{$\blacklozenge$} & \multirow{1}{0.7cm}{\checkmark} & \multirow{1}{0.7cm}{} & \multirow{1}{0.7cm}{\checkmark} & \multirow{1}{0.7cm}{} & \multirow{1}{0.7cm}{} & \multirow{1}{0.7cm}{}\\ \hline
              
                Terrizzano et al. (2015)&\cite{Terrizzano2015}& \multirow{1}{0.7cm}{$\blacklozenge$} & \multirow{1}{0.7cm}{\checkmark} & \multirow{1}{0.7cm}{\checkmark} & \multirow{1}{0.7cm}{} & \multirow{1}{0.7cm}{} & \multirow{1}{0.7cm}{\checkmark} & \multirow{1}{0.7cm}{\checkmark}\\ \hline
                
                Constance (Hai et al., 2016)&\cite{Hai2016}& \multirow{1}{0.7cm}{$\blacklozenge$} & \multirow{1}{0.7cm}{\checkmark} & \multirow{1}{0.7cm}{\checkmark} & \multirow{1}{0.7cm}{} & \multirow{1}{0.7cm}{} & \multirow{1}{0.7cm}{} & \multirow{1}{0.7cm}{}\\ \hline
                
                GEMMS (Quix et al., 2016)&\cite{Quix2016B}& \multirow{1}{0.7cm}{$\lozenge$} & \multirow{1}{0.7cm}{\checkmark} & \multirow{1}{0.7cm}{} & \multirow{1}{0.7cm}{} & \multirow{1}{0.7cm}{} & \multirow{1}{0.7cm}{} & \multirow{1}{0.7cm}{}\\ \hline
                
                CLAMS (Farid et al., 2016)&\cite{Farid2016} & \multirow{1}{0.7cm}{$\blacklozenge$} & \multirow{1}{0.7cm}{\checkmark} & \multirow{1}{0.7cm}{} & \multirow{1}{0.7cm}{} & \multirow{1}{0.7cm}{} & \multirow{1}{0.7cm}{} & \multirow{1}{0.7cm}{}\\ \hline
            
                Suriarachchi and Plale (2016)&\cite{Suriarachchi2016}& \multirow{1}{0.7cm}{$\lozenge$} & \multirow{1}{0.7cm}{} & \multirow{1}{0.7cm}{} & \multirow{1}{0.7cm}{} & \multirow{1}{0.7cm}{\checkmark} & \multirow{1}{0.7cm}{} & \multirow{1}{0.7cm}{\checkmark}\\ \hline
                
                Singh et al. (2016)&\cite{Singh2016}& \multirow{1}{0.7cm}{$\blacklozenge$} & \multirow{1}{0.7cm}{\checkmark} & \multirow{1}{0.7cm}{\checkmark} & \multirow{1}{0.7cm}{\checkmark} & \multirow{1}{0.7cm}{\checkmark} & \multirow{1}{0.7cm}{} & \multirow{1}{0.7cm}{}\\ \hline
                
                Farrugia et al. (2016)&\cite{Farrugia2016}& \multirow{1}{0.7cm}{$\blacklozenge$} & \multirow{1}{0.7cm}{} & \multirow{1}{0.7cm}{} & \multirow{1}{0.7cm}{\checkmark} & \multirow{1}{0.7cm}{} & \multirow{1}{0.7cm}{} & \multirow{1}{0.7cm}{}\\ \hline
                
                GOODS (Halevy et al., 2016)&\cite{Halevy2016}& \multirow{1}{0.7cm}{$\blacklozenge$} & \multirow{1}{0.7cm}{\checkmark} & \multirow{1}{0.7cm}{\checkmark} & \multirow{1}{0.7cm}{\checkmark} & \multirow{1}{0.7cm}{} & \multirow{1}{0.7cm}{\checkmark} & \multirow{1}{0.7cm}{\checkmark}\\ \hline
     
                CoreDB (Beheshti et al., 2017)&\cite{Beheshti2017}& \multirow{1}{0.7cm}{$\blacklozenge$} & \multirow{1}{0.7cm}{} & \multirow{1}{0.7cm}{\checkmark} & \multirow{1}{0.7cm}{} & \multirow{1}{0.7cm}{} & \multirow{1}{0.7cm}{} & \multirow{1}{0.7cm}{\checkmark}\\ \hline
     
                Ground (Hellerstein et al., 2017)&\cite{Hellerstein2017}& \multirow{1}{0.7cm}{$\lozenge \sharp$} & \multirow{1}{0.7cm}{\checkmark} & \multirow{1}{0.7cm}{\checkmark} & \multirow{1}{0.7cm}{} & \multirow{1}{0.7cm}{} & \multirow{1}{0.7cm}{\checkmark} & \multirow{1}{0.7cm}{\checkmark}\\ \hline
                
                KAYAK (Maccioni and Torlone, 2018)&\cite{Maccioni2018}& \multirow{1}{0.7cm}{$\blacklozenge$} & \multirow{1}{0.7cm}{\checkmark} & \multirow{1}{0.7cm}{\checkmark} & \multirow{1}{0.7cm}{\checkmark} & \multirow{1}{0.7cm}{} & \multirow{1}{0.7cm}{} & \multirow{1}{0.7cm}{}\\ \hline
                
                CoreKG (Beheshti et al., 2018)&\cite{Beheshti2018}& \multirow{1}{0.7cm}{$\blacklozenge$} & \multirow{1}{0.7cm}{\checkmark} & \multirow{1}{0.7cm}{\checkmark} & \multirow{1}{0.7cm}{\checkmark} & \multirow{1}{0.7cm}{\checkmark} & \multirow{1}{0.7cm}{} & \multirow{1}{0.7cm}{\checkmark}\\ \hline
                
                Diamantini et al. (2018)&\cite{Diamantini2018}& \multirow{1}{0.7cm}{$\lozenge$} & \multirow{1}{0.7cm}{\checkmark} & \multirow{1}{0.7cm}{} & \multirow{1}{0.7cm}{\checkmark} & \multirow{1}{0.7cm}{\checkmark} & \multirow{1}{0.7cm}{} & \multirow{1}{0.7cm}{}\\ \hline
            \end{tabular}
            
            \begin{flushleft}
                $\blacklozenge:~$Data lake implementation~~$\lozenge:~$Metadata model\\
                $\sharp \ :~$Model or implementation assimilable to a data lake
            \end{flushleft}

            \label{tab:comparaison}
        \end{table*}

        In terms of functionalities, we note an almost unanimous agreement on the relevance of semantic enrichment, with 12 out of 15 systems offering this feature and, to a lesser extent, of data indexing  (9/15) and data link generation (8/15). On the other hand, other features are much less shared, especially data polymorphism (4/15) and data versioning (3/15). In our opinion, this rarity does not indicate a lack of relevance, but rather implementation complexity. Such features are indeed mainly found in the most complete systems (GOODS, CoreKG and Ground) and can therefore be considered as advanced features. 

\section{Metadata Typology}
\label{sec:typ_metadata}

    The comparison results from Section~\ref{sec:comparaison_fonct} show that no metadata system offers all expected functionalities. To bridge this gap, we propose in the following a metadata model that supports all six key functionalities. Beforehand, we need a generic concept that represents any set of homogeneous data that the model can process. In the literature, we find data units \cite{Quix2016B}, entities \cite{Beheshti2017}, datasets \cite{Maccioni2018} and objects \cite{Diamantini2018}. We adopt objects, which seem more appropriate to represent a dataset in an abstract way. More precisely, an object may be a relational table or a physical file (spreadsheet document, XML or JSON document, text document, tweet collection, image, video, etc.).

    The definition of a metadata model for data lakes also involves identifying the metadata to be considered. To this end, we extend a medatata typology that categorizes metadata into intra-object, inter-object and global metadata~\cite{Sawadogo2019} with new types of inter-object (relationships) and global (index, event logs) metadata.

    \subsection{Intra-object Metadata}
    \label{sec:typ_intra}
    
        This category refers to metadata associated with a given object.
        
        \noindent\textbf{Properties} provide a general description of an object, in the form of key-value pairs. Such metadata are usually obtained from the filesystem: object title, size, date of last modification, access path, etc.
         
        \noindent\textbf{Summaries and previews} provide an overview of the content or structure of an object. They can take the form of a data schema in a structured or semi-structured data context, or a word cloud for textual data. 
        
        \noindent Raw data in the lake are often modified through updates that result in the creation of new \textbf{versions} of the initial data, which can be considered as metadata. Similarly, raw data (especially unstructured data) can be reformatted for a specific use, inducing the creation of new \textbf{representations} of an object. 
        
        \noindent\textbf{Semantic metadata} are annotations that help understand the meaning of data. More concretely, they are descriptive tags, textual descriptions or business categories. Semantic metadata are often used  for detecting object relationships.

    \subsection{Inter-object Metadata}
    \label{sec:typ_inter}
        
        Inter-object metadata account for relationships between at least two objects.
    
        \noindent\textbf{Objects groupings} organize objects into collections, each object being able to belong simultaneously to several collections. Such groups can be automatically derived from semantic metadata such as tags and business categories. Some properties can also be used for generating groups, e.g., objects can be grouped w.r.t. format or language. 

        \noindent\textbf{Similarity links} reflect the strength of the similarity between two objects. Unlike object groupings, similarity relationships refer to the intrinsic properties of objects, such as their content or structure. For example, it may be the common word rate between two textual documents, a measure of the compatibility of the schemas of two structured or semi-structured objects~\cite{Maccioni2018}, or other common similarity measures.
        
        \noindent\textbf{Parenthood relationships}, which we add to our reference typology \cite{Sawadogo2019}, reflect the fact that an object can be the result of joining several others. In such a case, there is a ``parenthood'' relationship between the combined objects and the resulting object, and a ``co-parenthood'' relationship between the merged objects. This type of relationship thus makes it possible to take advantage of the processing carried out in the data lake to identify objects that can be used together, in addition to maintaining traceability of the origin of the objects generated inside the lake. 
    
    \subsection{Global Metadata}
    \label{sec:typ_glob}
    
        Global metadata are data structures designed to provide a contextual layer to the lake's data, to facilitate and optimize its analysis. Unlike intra and inter-object metadata, global metadata potentially concern the entire data lake. In addition to the semantic resources identified in our reference typology \cite{Sawadogo2019}, we propose two new types of global metadata.
        
        \noindent\textbf{Semantic resources} are essentially knowledge bases (ontologies, taxonomies, thesauri, dictionaries) used to generate other metadata and improve analyses. For example, a thesaurus can help extend a query by associating synonyms of the terms typed by the user. Similarly, a thesaurus can be used while generating object groupings, to merge collections from different but equivalent tags. 
        
        Semantic resources are generally coming from external sources. This is typically the case for ontologies that are provided by knowledge bases on the Internet. However, in some cases, semantic resources can be created and customized specifically for the management and analysis of lake data. For instance, a business ontology can thus be used to define abstract tags allowing to group together several equivalent or close tags during analysis.
        
        \noindent\textbf{Indexes}
        are data structures that help find an object quickly. They establish (or measure) the correspondence between characteristics such as keywords, patterns or colors, with the objects contained in the data lake. Indexes can be simple (textual indexing) or more complex (e.g., on images or sound content). They are mainly used to search for data in the lake.

        \noindent\textbf{Logs} are used to track user interactions with the data lake. This involves the sequential recording of events such as users logging in, viewing or modifying an object. Such metadata help analyze data lake usage by identifying the most consulted objects or studying user behaviour.

    \subsection{Formal Definition of a Data Lake}

        From the above typology, we can now formally define a data lake.    
        
        \begin{definition}
            A data lake is a pair $DL = \langle \mathcal{D}, \mathcal{M} \rangle$, where $\mathcal{D}$ is a set of raw data and $\mathcal{M}$ a set of metadata describing the objects of $\mathcal{D}$. Objects in $\mathcal{D}$ can take the form of structured (relational database tables, CSV files, etc.), semi-structured (JSON, XML, YAML documents, etc.) and unstructured data (images, textual documents, videos, etc.). Metadata are subdivided into three components: $\mathcal{M} = \langle \mathcal{M}_{intra}, \mathcal{M}_{inter}, \mathcal{M}_{glob} \rangle$, where $\mathcal{M}_{intra}$ is the set of intra-object metadata, $\mathcal{M}_{inter}$ the set of inter-object metadata and $\mathcal{M}_{glob}$ the set of global metadata.
        \end{definition}

\section{Metadata Model}
\label{sec:repres_concep}

        MEDAL adopts a logical metadata representation based on the hypergraph, nested graph and attributed graph notions. We represent an object by an \textbf{hypernode} containing various elements (versions and representations, properties, etc.). Hypernodes can be linked together (similarity, parenthood, etc.).
        
    \subsection{Intra-object Metadata}
    \label{sec:modele-meta-intra}

        Each hypernode contains \textbf{representations}, reflecting the fact that data associated with an object can be presented in different ways. There is at least one representation per hypernode, corresponding to raw data. 
        Other representations all derive from this initial representation. Each representation corresponds to a node bearing attributes, simple or complex. These are the properties of the representation. A representation can be associated with an object actually stored in the lake or be a view calculated on demand. 
        
        The transition from one representation to another is done via a \textbf{transformation}. It takes the form of a directed edge connecting two representation nodes. This edge also bears attributes, which are the properties describing the transformation process 
        from the first representation to the second (full script or description, in case of manual transformation). 
                
        A hypernode can also contain \textbf{versions}, which are used to manage the evolution of lake data over time. We also associate versions with nodes bearing attributes. 
        The creation of a new version node is not necessarily systematic at the slightest change. Depending on the nature and frequency of data evolution, it is possible to implement various strategies, such as those used to manage slowly changing dimensions in data warehouses~\cite{Kimball2008}.
        The creation of a new version is done via an \textbf{update} similar to a transformation, since it is also translated by a directed edge and possesses some attributes.
        
        Finally, a hypernode also bears attributes such as the origin of the object or aggregates of the attributes of the representations and versions it contains (number of versions, representations, total size, etc.).     
        Thus, a hypernode contains a tree whose nodes are representations or versions and directed edges are transformations or updates. One representation (resp. version) is derived from another by a transformation (resp. update). A version can lead to a representation via a transformation, but a version cannot be derived from a representation. Thus, the root of the tree is the initial raw representation of the hypernode and each version has its own subtree of representations. 
        
        \begin{definition}
            Let $\mathcal{N}$ be a set of nodes. The set of \emph{intra-object metadata} $\mathcal{M}_{intra}$ is the set of hypernodes such that $\forall h \in \mathcal{M}_{intra}, h = \langle N, E \rangle$, where 
                $N \subset \mathcal{N}$ is the set of nodes (representations and versions) carrying attributes of $h$ and
                $E = \{ r_{(transformation\ |\ update)} \in N \times N\}$ is the set of edges (transformations and updates) carrying attributes of $h$.
        \end{definition}
        
\begin{figure*}[hbt]
            \centering
            \begin{minipage}{.5\textwidth}
                \centering
                \includegraphics[width=.5\linewidth]{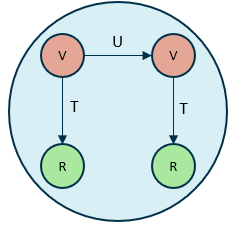}
                \captionsetup{justification=centering}
                \captionof{figure}{Sample hypernode and its \newline representation tree}
                \label{fig:intra}
            \end{minipage}%
            \begin{minipage}{.5\textwidth}
                \centering
                \includegraphics[width=\linewidth]{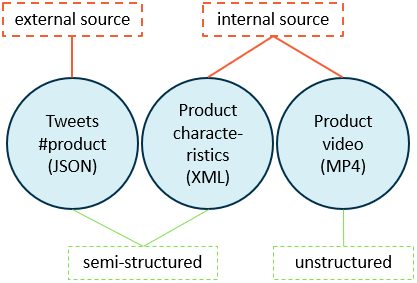} 
                \captionsetup{justification=centering}
                \captionof{figure}{Sample interconnected hypernodes}
                \label{fig:inter}
            \end{minipage}
        \end{figure*}

        Let us illustrate these notions with an example (Figure~\ref{fig:intra}). Imagine a company selling various products. Information on these products (name, unit price, description, etc.) is stored in the lake as an XML file. A hypernode describes this dataset and has a version node that corresponds to the initially ingested XML file. To assist in querying product information, a user decides to extract the XML file's schema. This generates a new representation. Now suppose that the price of some products changes and new products are added to the catalogue. This change in data generates a new version, linked to the first version by an update. Finally, if the user wants to obtain the schema of the most recent data, this creates a new representation coming from the second version.

        \vspace*{-0.3cm}
    
    \subsection{Inter-object Metadata}
    \label{sec:modele-meta-inter}

        An object grouping is modeled by a set of non-oriented hyperedges, i.e., edges that can link more than two (hyper)nodes. Each hyperedge corresponds to a collection of objects. If grouping is performed on a hypernode attribute, a hypernode belongs to the hyperarc that corresponds to its value for the attribute. Thus, there are as many hyperarcs as there are distinct values of the considered attribute. Note that not all attributes are necessarily used in groupings and that groupings can be made on other elements but attributes. 

        A similarity link between two hypernodes is represented by a non-oriented edge with attributes: value of the similarity metric, type of metric used, date of the metric, etc. Two hypernodes connected by a similarity link must be comparable, i.e., they must each contain a representation that can be compared to the other with a similarity measure. 
        
        A hypernode can be derived from other hypernodes through a parenthood link. To translate this relationship, we use an oriented hyperedge: all the ``parent'' hypernodes and the ``child'' hypernode are connected by this oriented hyperedge toward the child hypernode. This hyperedge also bears descriptive attributes.
        
        \begin{definition}
            The set of \emph{inter-object metadata} $\mathcal{M}_{inter}$ is defined by three pairs $\langle H, E_g \rangle$, $\langle H', E_s \rangle$ and $\langle H'', E_p \rangle$, where
             $H \subset \mathcal{M}_{intra}$, $H' \subset \mathcal{M}_{intra}$ and $H'' \subset \mathcal{M}_{intra}$ are sets of hypernodes carrying attributes;
             $E_g = \{ E_g^{param} \ | \ E_g^{param} : H \rightarrow \mathcal{P}(H)\}$ is the set of functions grouping hypernodes in collections w.r.t. a given parameter (often an attribute);
             $E_s = \{ s \ | \ s \in H' \times H'\}$ is the set of edges (similarity links) carrying attributes; and
             $E_p = \{ (h_1, ..., h_n, h_{child}) \ | \ (h_1, ..., h_n, h_{child}) \in (H'')^{n+1}\}$ is the set of parenthood relationships, with $(h_1, ..., h_n)$ being the parent hypernodes ($n \geq 2$) and $h_{child}$ the child hypernode.

        \end{definition}

        Let us pursue the example of Section~\ref{sec:modele-meta-intra} by adding other hypernodes: tweets related to the company and a commercial video of the products. In a grouping on the origin of data, the tweet hypernode is alone in the ``external source'' collection, while the other two are in the ``internal source'' collection. In a second grouping on the format of the initial version, the video hypernode is alone in the ``unstructured'' collection, while the other two hypernodes are in the ``semi-structured'' collection. Collections are represented by dotted rectangles in Figure~\ref{fig:inter} (attributes of the hypernodes are not represented for simplicity).

    \subsection{Global Metadata}
    
        Global metadata are specific elements that are managed differently from other metadata. They ``gravitate'' around hypernodes and are exploited as needed, i.e., almost systematically, especially logs and indexes. We consider that semantic resources are stored in nodes, while indexes and event logs are rather physical structures and are highly dependent on the technology used to implement the data lake and the metadata system. 

\section{Conclusion}
\label{sec:conclusion}

    After an overview of the definitions of a data lake from the literature, we propose in this paper our own definition of this concept. Then, we identify the six key features that the metadata system of a data lake must provide to be as 
    robust as possible in addressing the big data issues and the schema-on-read approach. 
    Comparing existing metadata systems, we show that some succeed in providing most features, but none offers them all.
    
    Hence, we propose a new metadata model, MEDAL, based on the notion of object and a typology of metadata in three categories: intra-object, inter-object and global metadata. 
    MEDAL adopts a graph-based organization. An object is represented by a hypernode containing nodes that correspond to the versions and representations of an object. Transformation and update operations are modeled by oriented edges linking the nodes. Hypernodes can be linked in several ways: edges to model similarity links and hyperarcs to translate parenthood relationships and object groupings. Finally, global resources are also present, in the form of knowledge bases, indexes or event logs. 
   
    Thanks to all these elements, MEDAL supports all six key features we have identified, making it the most complete metadata model to the best of our knowledge. However, MEDAL is not implemented yet. It is the objective of future work in which we shall propose an application of our metadata model in a context of structured, semi-structured and unstructured data. This implementation will allow us to evaluate MEDAL in more detail, in particular by comparing it with other existing systems.

\section*{Acknowledgments}

    Part of the research presented in this article is funded by the Auvergne-Rhône-Alpes Region, as part of the AURA-PMI project that finances Pegdwendé Nicolas Sawadogo's PhD thesis.





\end{document}